(RESEARCH ARTICLE)

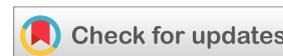

# Redefining Computing: Rise of ARM from consumer to Cloud for energy efficiency

Tahmid Noor Rahman *, Nusaiba Khan and Zarif Ishmam Zaman

*Department of EEE, BSRM School of Engineering, BRAC University, Merul Badda, Dhaka-1212, Bangladesh.*

## Abstract

Today, our lifestyle revolves around digital devices powered by microprocessors of different instruction set architectures (ISA). Among them, the most common are x86 and ARM, the brainpower of our computers and smartphones. Reduced instruction set computing (RISC) is the basis of ARM architecture, designed to offer greater energy efficiency. On the other hand, principles of complex instruction set computing (CISC) are utilized by x86 processors, which handle heavier computing tasks while being more power-hungry. The rise of smartphones over a decade has changed the laptop market. It also influenced customers towards ARM-based energy-efficient laptops. This transition in the computing segment is seen not only in consumers but also in commercial settings, especially data centers. Usually, data centers are designed to operate 24/7, where energy is a big concern. So, ARM chips have started making their way to cloud servers. This paper comprehensively analyzes and compares the ARM and x86 architectures to unravel the factors contributing to ARM's increasing dominance in the market. It also explores the impact of smartphones on the laptop market and assesses the significance of system-on-a-chip (SoC). Focusing on the proper utilization of energy and sustainability, our paper offers valuable insights into the growing trend of adopting ARM processors in the computing industry.

## 1. Introduction

ARM and x86 are two processor architectures widely used in various computing devices, such as smartphones, tablets, laptops, desktops, servers, and supercomputers. Processor architectures are the sets of instructions the processor can execute, affecting how the processor performs different tasks and interacts with other components.

ISA (Instruction Set Architecture) is the part of a computer processor that defines its machine language instructions. An instruction set provides the list of commands a processor understands and executes. It acts like a translator, converting assembly/machine code written by programmers into signals the processor can process. The ISA is the boundary between a computer's software (applications, operating systems) and its physical hardware (CPU and components).

RISC (Reduced Instruction Set Computing) uses a smaller, simplified set of instructions that take less time to execute. This allows RISC processors to run instructions faster and be more efficient. CISC (Complex Instruction Set Computing) has a more extensive, complex set of instructions. These instructions take more time to execute but allow more to be done in a single instruction. RISC prioritizes simplicity and speed, while CISC prioritizes complexity and capability per instruction.

System-on-a-chip (SoC) represents a groundbreaking innovation in microelectronics, revolutionizing the integration of diverse electronic functions onto a single semiconductor chip. The first SoC solution evolved from the Hamilton Pulsar "Wrist Computer" digital watch unveiled on the Johnny Carson Show in 1970. The watch was designed by George Thiess and Willy Crabtree at Electro-Data, Inc., and contained 44 chips and 4,000 bonding wires. It was notoriously unreliable until RCA engineers reduced the timekeeping circuitry to one chip. [1]

---

* Corresponding author: Tahmid Noor Rahman

The first RISC processor was the IBM 801, designed by John Cocke in 1980. The IBM 801 was never commercialized, but it influenced a generation of RISC designs, such as the MIPS, the SPARC, and the ARM. The first CISC processor was the Intel 4004, designed by Federico Faggin, Ted Hoff, and Stanley Mazor in 1971. Thus, the first device to utilize the Intel 4004 microprocessor was the Busicom 141-PF printing calculator, released in 1971. [2] [3]

ARM and x86 have evolved and improved over the years, adding new features and extensions to support the changing demands and trends of the computing industry. For example, both architectures have added support for 64-bit operations, which enable the processors to work efficiently with giant data sets and massive amounts of RAM.

## 2. Evolution

The first ARM chip, the Acorn RISC Machine (later renamed Advanced RISC Machine), was developed by the advanced research and development team of Acorn Computers, a British microcomputer pioneer. Acorn was one of the leading names in the British personal computer market. The development of what was to become ARM began in 1983. The world's first commercial RISC and ARM processor, ARM1, was fabricated in April 1985 at VLSI Technology. [4]

These samples were manufactured using a three μm process. The experience of designing ARM1 and programming sample chips has shown that there are some areas where the set of instructions could be improved to maximize the performance of the systems based on it. This extension facilitated the processing of digital signals in real-time, which was to be used to generate sounds, an essential feature of family and educational computers. Later developed ARM2 maintained its small cutting size and low number of transistors with all these additions. [4]

In 1987, Archimedes was released as the first commercial for ARM, which included an 8MHz version of ARM2 and three MEMC, VIDC, and IOC support chips, input and output controllers, and simple operating systems. It took two to three years to develop a credible number of ARM and Archimedes native applications software. Since then, Acorn has refined and improved its computer models and confirmed its position as a British home and educational computer market leader. After the launch of Archimedes, Acorn continued to support the research and development team in creating improved versions of the chip, which provided a better performance. 1989 ARM3 was launched at a significantly higher clock speed of 25MHz. Acorn's desktop computers using this chip were launched in 1990. [4]

Interest in the ARM family grew as more designers started working on RISC, and the design of ARM was seen to match a defined need for high-performance, low-power, low-cost RISC processors. Acorn RISC Machine became Advance RISC Machine, and the company Advances RICS Machines Ltd. was founded. ARM Ltd. was established with a clear mission to continue the development of the ARM processor. It is now under the umbrella of the Japanese investment firm SoftBank Group and called ARM Holdings. The first development of ARM Ltd was the next step of the ARM3 processor, called ARM6. [4]

From ARM1 to ARM6, all were 32-bit processors with 26-bit address bus. This limitation meant they could only directly access up to 64 MB of main memory. Even though each processor was better than its predecessor, support for 32-bit buses was added with ARM7, released in 1994. [4]

Intel has played a vital role in developing the complex instruction set computing (CISC) architecture. Intel's first x86 microprocessor was the Intel 8086, launched in 1978. Intel designed this to extend their existing 8-bit 8080 and 8085 processors. It had 29,000 transistors fabricated on a three μm process. [5]

Intel's 1982 release of the 80286 processor represented the company's first 16-bit chip in the x86 family, with a dedicated memory protection unit and 24-bit addressing, expanding maximum memory from 1MB (8086) to 16MB. It operated at up to 12.5 MHz, 2-3X faster than the 8086. [5]

The 80286 was crucial in bringing Intel's x86 family from 8-bit to full 32-bit computing. It enabled the transition to this 16-bit era for the final release of the 80386 in 1985 as Intel's first fully 32-bit processor capable of handling up to 4GB of memory. The 80386 represented, in particular, the foundation of Intel's first era of dominance, as its 1986 edition with 275,000 transistors, 32-bit operation, and processing speeds up to 33MHz outnumbered rivals by a considerable margin to become a cornerstone of computing for the next decade across desktops and servers. [5]

The x86-64 or simply x64 architecture is a 64-bit version of the x86 instruction set. Now, every computer uses x64, though they are called x86 by the legacy of the architecture. The main difference between x86 and x64 is the data they can handle with each clock cycle and the processor's register width. A register is a small and fast memory unit inside the CPU that stores data temporarily. x86 has eight general-purpose registers, each 32 bits wide, limiting the system to a maximum of 4GB RAM, while x64 has 16 general-purpose registers, each 64 bits wide. This update allows the x64 system to address up to 18.4 million terabytes of RAM theoretically. An essential aspect of the x64 architecture is its compatibility. While the x64 system can run 32-bit software, it also offers the advantage of increasing performance and using more significant amounts of RAM. [6]

- In 1995, AMD started R&D of x86-64 as an evolutionary step from x86. The goal is to address memory limitations while maintaining compatibility.
- In 1999, AMD publicly unveiled and released the first x86-64 processor architecture and Athlon 64 chips supporting 64-bit instruction extensions.
- In early 2000, AMD Athlon 64 processors started being popularized and drove the adoption of 64-bit computing in mainstream PCs. Backward compatibility facilitates migration.
- In mid-2000, Intel introduced its IA-32e architecture to enable 64-bit extensions in Pentium 4 chips. This helps solidify x64 as the successor to x86 on both AMD and Intel platforms.

ARM64, or AArch64, is a 64-bit architecture developed by ARM Holdings. ARM64 is designed to operate with 64-bit memory addresses, enabling larger memory spaces than its 32-bit counterpart (ARMv7). These processors are also backwards compatible with 32-bit ARMv7 software like x64 and x86 in CISC. [7]

- In 2011, ARMv8-A architecture, including the ARM64 instruction set, was announced by ARM Holdings. It was designed to bring 64-bit processing capabilities to ARM-based processors.
- In 2012, ARMv8-A specifications were released to ARM partners, enabling them to design and produce ARM64 processors.
- In 2013, The first ARMv8-A architecture-based processors, including ARM64 designs, began to appear.

## 3. Literature Review

Historically, debates around RISC and CISC architectures have subsided, yet the x86 CISC architecture remains prevalent. Isen et al. highlighted that, despite translation complexities, CISC processors, particularly x86, can approach RISC performance through micro-architectural techniques. This underscores the resilience of x86 architecture and its adaptability to contemporary demands. [8]

The perspective of Bhandarkar et al. on RISC and CISC arguments resurfaces, emphasizing the enduring advantages of RISC architectures. Their study compared MIPS M/2000 (RISC) and Digital VAX 8700 (CISC) on SPEC benchmarks, revealing a consistent net advantage for RISC regarding cycles per program. This advantage ranges from slightly under a factor of 2 to almost a factor of 4, indicating the architectural prowess of RISC. [9]

A comparison by Abudaga et al. evaluated the performance of ARM and x86 microprocessors, considering both ordered and non-order CPU models. The results highlight the superiority of ARM in various computational scenarios and emphasize its potential as a versatile and powerful alternative to x86. [10]

ARM processors have undoubtedly dominated the mobile device market, showing a remarkable computing and energy ratio. The recent entry into server-grade processors, such as the AMD A1100 series, has made the ARM64 a powerful player in cloud data centers and extensive data analysis. Kalyanasundaram et al. suggested that comparative studies have shown the competitive performance and energy efficiency of the ARM64 server compared to the x64 server, which means a potential reduction in the operational cost of the data center. [11]

Evaluation of ARM-based computers for quantum chemistry applications shows subtle performance differences. ARM64 CPUs demonstrate energy efficiency advantages over x86 CPUs, especially for memory-intensive calculations. This view of Keipert et al. expanded the scope of ARM processors and positioned them as competitive choices for specific computational workloads. [12]

A study by Otto et al. analyzes the power, performance, and energy efficiency of Harris Corner detection computer vision algorithms on embedded platforms with ARM and x86 multicore processors, with real-time operation targeting low power levels of milliwatts. It reveals that systematic cross-layer optimization at the application, compiler, and system levels significantly improves the energy efficiency of corner detection while meeting real-time performance constraints. This optimization improves Harris Corner energy efficiency on Atom and ARM by 89.5% and 87.2%, respectively, compared to the standard ARM with Thumb ISA and Intel Atom with x86 ISA. [13]

Exploring the energy and cost-effectiveness of ARM processor-based clusters versus Intel x86 workstations provides insight into various applications. Ou et al. highlighted scenarios where ARM clusters have superior energy efficiency, especially in light-based computational applications. However, it recognizes the diminishing advantages of computationally intensive tasks. [14]

In response to the growing importance of green computing, Chada et al.'s study introduced statistical power estimation techniques using performance monitoring meters (PMC) for x86 systems. The application of statistical methods to ARM systems is parallel and emphasizes the importance of accurate and reliable power models for energy-conscious performance optimization in all architectures. [15]

The versatility of ARM architectures is underlined in their transition from experiments to omnipresence in modern devices. Gupta et al. found that ARM prioritizes low power consumption and considerable performance, unlike x86, designed for high performance. Recent advances position ARM processors as more potent than x86 processors, leading to changes in industrial dynamics and influencing software development environments. [16]

ARM's growth path transcends its mobile roots, exploring server-grade components focusing on scalability. The availability of mobile and server-class ARM SoCs demonstrates ARM's adaptability to various computing needs. The study of scaling architectures by Azimi et al. placed ARM as a competitor for high-performance computing (HPC), including scientific applications and workloads of artificial intelligence. [17]

Experimental comparisons between x86 and ARM architectures explore the feasibility of building low-power server architectures. Analysis of Aroka et al. included web and database servers, evaluating power use, CPU load, temperature, request delays, and several handled requests. ARM systems are a prudent choice when energy efficiency is at the top without compromising performance. [18]

Looking at the performance of the ARM A64FX in supercomputing, it seems capable of challenging the domination of the x86-64 processors. A64FX's performance varies depending on how it stacks against Intel and AMD processors. This shows how ARM influences high-performance computing and its evolving role, as suggested by Kuzminsky et al. [19]

The delicate balance between performance and power efficiency remains an important area of research, with ARM and RISC architectures ready to play an essential role in shaping the future of computing. As the industry evolves, these buildings are beacons of energy-conscious computer use.

## 4. Phones

A phone is a communication device designed to transmit voice or other data types over a distance. The evolution of the phone has seen significant technological advancements, transforming from wired telephones to the modern wireless mobile phones we use today. The history of the phone dates back to the 19th century when Alexander Graham Bell patented the invention of the telephone in 1876. Early phones were wired devices that required a physical connection between two locations. The idea of cellular mobile wireless phones emerged in the 1970s and became a reality in the 1980s. Instead of wires, radio signals and cellular tower networks would enable phone calls from anywhere. Motorola launched the first commercial mobile phone in 1983, the Motorola DynaTAC 8000x. [20]

### 4.1. Smart + Phone = Smartphone

The term "smartphone" is a combination of the words "smart" and "phone," referring to phones that have become smarter over time by integrating advanced computing technology and software. A smartphone is a mobile device that combines the functions of a mobile phone with those of a personal digital assistant (PDA) and, in more modern iterations, a handheld computer. Early PDAs such as Casio Digital Diary SF-R20 (1984) and Sharp Wizard (1989) were intended to be mobile organizers with calendars/contacts, note-taking, and some basic applications. However, they lacked cellular phone capabilities. As mobile phones became smaller and gained basic PDA functions in the late 1990s, like contacts and calendars, they were not considered smartphones since they ran minimal software. Phones were still mainly used for calling, texting, and basic utilities. The world's first smartphone is widely considered to be the IBM Simon Personal Communicator (SPC), also known as the IBM Simon or simply the Simon. It was introduced by IBM, manufactured by Mitsubishi Electric, and debuted on August 16, 1994. [21]

Actual smartphones emerged in the early 2000s as devices that integrated computer-like capabilities and dedicated operating systems, allowing third-party application development and installation.

### 4.2. CPU to SoC

The System on Chip (SoC) has become the unsung hero of the smartphone revolution. It has been a critical component that has fueled the advancement of smartphones over the last 15 years.

Early smartphones in the 90s relied on separate central processing units (CPUs) and components on the main board. However, achieving mass adoption's size, efficiency, and pricing was challenging.

In the mid-2000s, the first generation of smartphone SoCs emerged from companies like Texas Instruments, Qualcomm, and others. Although the title "first mobile SoC" is not formally awarded to any device or processor, the OMAP (Open Multimedia Applications Platform) series of Texas Instruments deserves vital consideration as a pioneer in this field. The OMAP 310 and 320 processors, launched in 2003, were among the first SoC solutions commercially available for mobile devices. Previously, phones were usually based on a patchwork of individual processors, which led to heavier devices and less efficient performance. The OMAP chip does not just combine the CPU into a single chip. They integrated vital functions such as memory controllers, sound cards, power management circuits, and graphics processing units (GPUs) into the same machine. This level of integration significantly improved power efficiency, reduced device size, and opened the way for a more robust and functional smartphone. [22]

---

## 5. Computers

In the 19th century, inventors such as Charles Babbage designed mechanical devices that could handle complex calculations. However, electronic computers emerged in the 20th century. The ENIAC (Electronic Numerical Integrator and Computer) was completed in 1945 and is often considered the world's first general-purpose electronic computer. It used massive vacuum tubes to calculate, covering a whole room. [23]

The 1970s witnessed a revolutionary development with the advent of microprocessors. The Altair 8800, released in 1975, became one of the first commercially successful microcomputer kits. Subsequently, IBM's introduction of the IBM PC 1981 brought personal computers (PCs) into homes and businesses. [24] [25]

### 5.1. Laptop - The All-in-One System

A laptop is a portable personal computer designed for mobility and convenience. It is also known as a notebook computer. Laptops are smaller and lighter than traditional desktop computers, intended to be carried easily for use in different locations. The concept of portable computing dates back to the 1970s when efforts were made to create computers that could be taken. The 1980s marked a significant period for mobile computing. Released in 1981, the Osborne 1 became the first commercially successful portable computer, featuring a built-in screen and floppy disk drives. However, the IBM PC Convertible, released in 1986, is often regarded as the first actual laptop. [26] [27]

### 5.2. Work on the go

The laptop revolutionized the modern office environment by providing unparalleled flexibility and mobility for professionals. Working from various locations, whether in the office, at home, or on a mobile basis, has become a cornerstone of modern work practices. Seamless professional transitions between different environments facilitate this flexibility without compromising productivity. Integrating collaboration tools and Internet connectivity into office laptops transformed remote collaboration, allowing virtual meetings, real-time document sharing, and effective project management.

Laptops are also crucial to facilitating flexible and dynamic learning experiences in educational environments. The portability of laptops allows students to engage in learning activities anywhere on campus, from classrooms to libraries. Students can use online textbooks, research databases, and educational software to improve access to educational resources significantly. Laptops foster student cooperation and enable them to work on group projects, share documents, and participate in virtual discussions. The multimedia functions of notebooks allow students to explore creative endeavors such as creating presentations, editing videos, and designing graphics.

## 6. Laptop or Smartphone

As discussed earlier, smartphones now have many basic computing capabilities that laptops were previously uniquely known for. In the past, when mobiles could not play high-resolution videos or browse the Internet effectively, laptops were the primary devices with such features. Internet connectivity while on the go posed a significant challenge. Despite the existence of Wi-Fi, it was limited to specific areas. Technologies like USB Modems and WiMax were introduced, allowing laptops to connect globally anytime, anywhere.

Over time, the rapid advancement of smartphones matched and now often exceeds laptop-level functionality in a device that fits in your pocket. Thanks to always-on cellular data connections that have evolved from 2G to 5G speeds. The graphics processing units (GPU) and digital signal processors (DSP) are now implemented in smartphone SoCs to make them more capable of graphics-intensive work like playing games or high-resolution videos.

In the past, one needed a desktop or laptop with an internet connection to communicate over the Internet. This scenario has now been seamlessly overtaken by smartphones, which can connect users to the rest of the world. These handheld devices, no more significant than the size of our palm, allow us to chat, send emails, watch videos, and even create high-resolution videos. Their multifunctionality and compact design have redefined how we interact with technology daily.

Global smartphone shipments are expected to reach 1.4 billion units in 2023, up from 1.2 billion in 2017, marking a significant 17% increase. International laptop shipments are expected to total around 200 million units in 2023, down from 269 million in 2017, representing a 26% decline. [28] [29]

## 7. The swap of ISAs

Traditionally, x86 processors have dominated the desktop and laptop market, while ARM processors have dominated the smartphone and tablet market. This is because x86 processors offer higher performance and compatibility with legacy software, while ARM processors offer lower power consumption and cost. However, there has been a swap of ISAs between these two markets in recent years, as ARM-based laptops and x86-based smartphones have emerged.

### 7.1. ARM Laptops

ARM-based laptops use processors based on the ARM ISA instead of the x86 ISA. These laptops are also known as Windows on ARM (WoA) devices because they run versions of Windows 10 compatible with ARM processors. The main advantage of ARM-based laptops is that ARM processors are more energy-efficient, require less cooling, offer longer battery life, and have a thinner design than x86-based laptops. ARM-based laptops also have standard smartphone features such as instant wake-up, mobile connectivity, and biometric authentication.

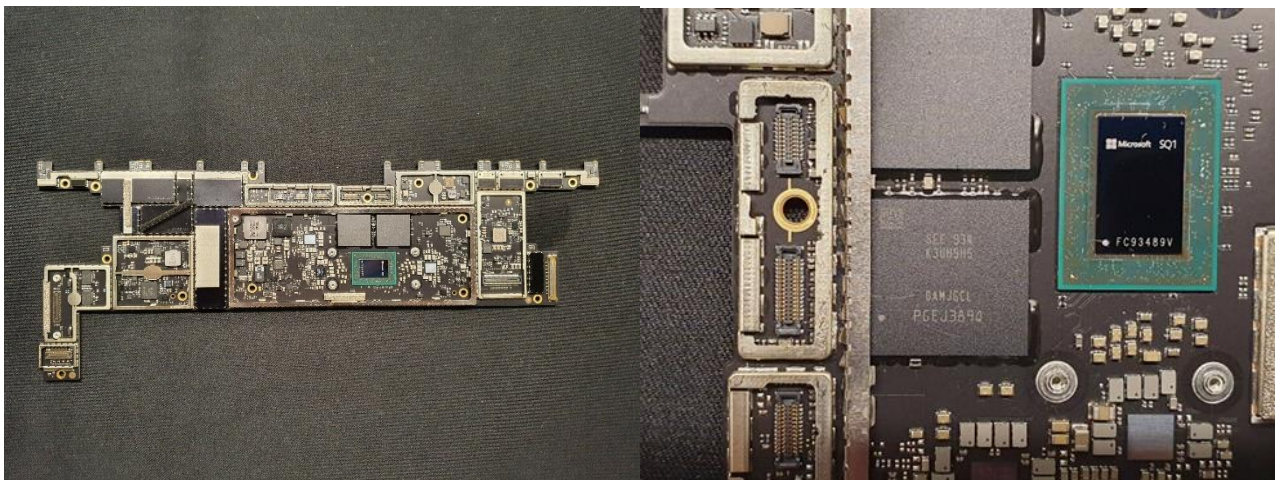

**Figure 1** PCB of Surface Pro X (Left); The customized ARM processor (Right)

**Image credit:** eBay

However, ARM-based laptops have disadvantages like low performance and compatibility problems. ARM processors are more potent than x86 processors, especially for tasks that require high computational intensity or graphic processing. In addition, ARM processors cannot run native x86 applications comprising most Windows software. To overcome this limitation, WoA devices use emulation and translation techniques to run x86 applications, but these methods can be significantly costly and do not support all applications.

One example of an ARM-based laptop is the Surface Pro X, a 2-in-1 device that was released by Microsoft in 2019. It uses a custom ARM processor called the Microsoft SQ1, co-developed by Microsoft and Qualcomm. The Microsoft SQ1 has a clock speed of 3 GHz and an integrated Adreno 685 GPU, making it one of the fastest ARM processors available. It has a battery life of up to 13 hours and supports LTE connectivity and Windows Hello facial recognition. This device runs Windows 10 Home on ARM and can run x86 applications using the Windows 10 emulation layer. However, the Pro X cannot run 64-bit x86 applications; it can only run 32-bit ones. [30]

### 7.2. x86 phones

x86-based smartphones use processors based on the x86 ISA rather than the ARM ISA. These smartphones are also known as Intel-powered smartphones, as they use processors made by Intel, the leading manufacturer of x86 processors. The main advantage of x86-based smartphones is that they offer higher performance and compatibility than ARM-based smartphones, as x86 processors have a higher clock speed and can run native x86 applications.

However, x86-based smartphones have drawbacks, such as higher power consumption and cost. x86 processors are less power efficient than ARM processors and require more battery capacity. x86 processors also have a higher price than ARM processors, as they are more complex and have higher licensing fees. Therefore, x86-based smartphones are best suited for users who value performance and compatibility over battery life and cost and mainly use x86 applications or want to use their smartphones as PCs.

One example of the x86-based smartphone is the Asus Zenfone 2, a flagship device released by Asus in 2015. Zenfone 2 uses an Intel Atom Z3580 processor with a clock speed of 2.3 GHz and an integrated PowerVR G6430 GPU. This phone has a battery capacity of 3000 mAh and supports fast charging and dual-SIM. It runs Android 5.0 Lollipop and can run ARM applications on x86 using the Intel Houdini binary translation layer. [31]

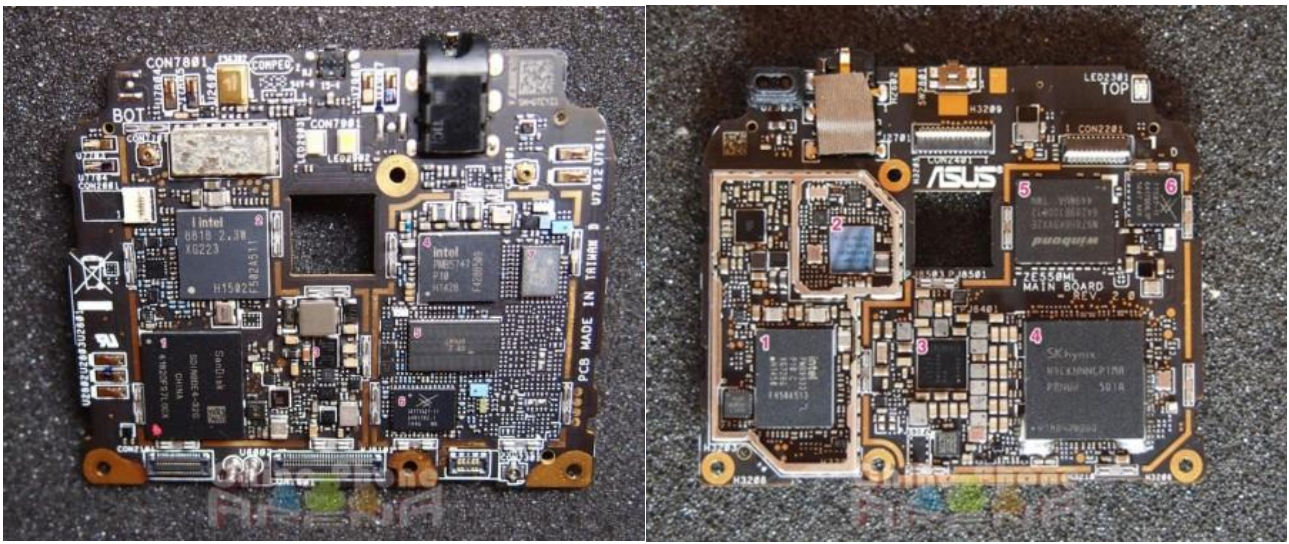

**Figure 2** PCB of Zenfone 2 Front, having Intel chips (Left); PCB of Zenfone 2 Back, having Intel chips (Right)

**Image credit:** China Phone Arena

### 7.3. Current Scenario

In the world of computers, the switch between different types of architectures for laptops and smartphones isn't a complete turnaround. Each type still has its good and not-so-good points in its own space.

ARM laptops are doing well because they offer longer battery life, a sleek design, and the ability to connect to cellular networks. These features are attractive to people who want a portable and mobile device. ARM laptops also benefit from the widespread use of ARM devices and software, especially in Android and Linux. There's an expectation that ARM laptops will get even better as more developers tweak their apps for ARM and as Microsoft makes it possible to run 64-bit x86 apps through emulation.

On the flip side, x86-based phones have faded away. They couldn't compete with ARM phones when it comes to being energy-efficient, cost-effective, and popular. One of the biggest problems for x86 phones was that most Android apps were designed for ARM processors, so they didn't work well on x86 ones. This lack of support held them back. Plus, x86 phones had a tough time gaining traction in a smartphone world mostly filled with ARM-based devices. They ended up being limited to niche markets and specific regions, never catching on widely.

Despite a 15% decline in PC shipments in 2022, with further drops expected in 2023, there was a rise in sales for ARM-based notebooks, according to Counterpoint Research. Apple dominated this market in 2022 with a 90% share. Counterpoint Research anticipates a sustained demand for ARM-powered laptops in the coming quarters, attributing this to the success of Apple's MacBooks, narrowing performance gaps with x86 CPUs, and vital ecosystem support.

While Apple has succeeded with its ARM-based Macs, most users still rely on Windows computers. The introduction of new ARM-powered SoCs for Windows by companies like MediaTek and Qualcomm in 2024 is expected to expedite the shift from x86 CPUs to ARM in mobile computers. Consequently, Counterpoint predicts that the share of ARM-based laptops will reach 21% in 2025 and 25% in 2027. [32]

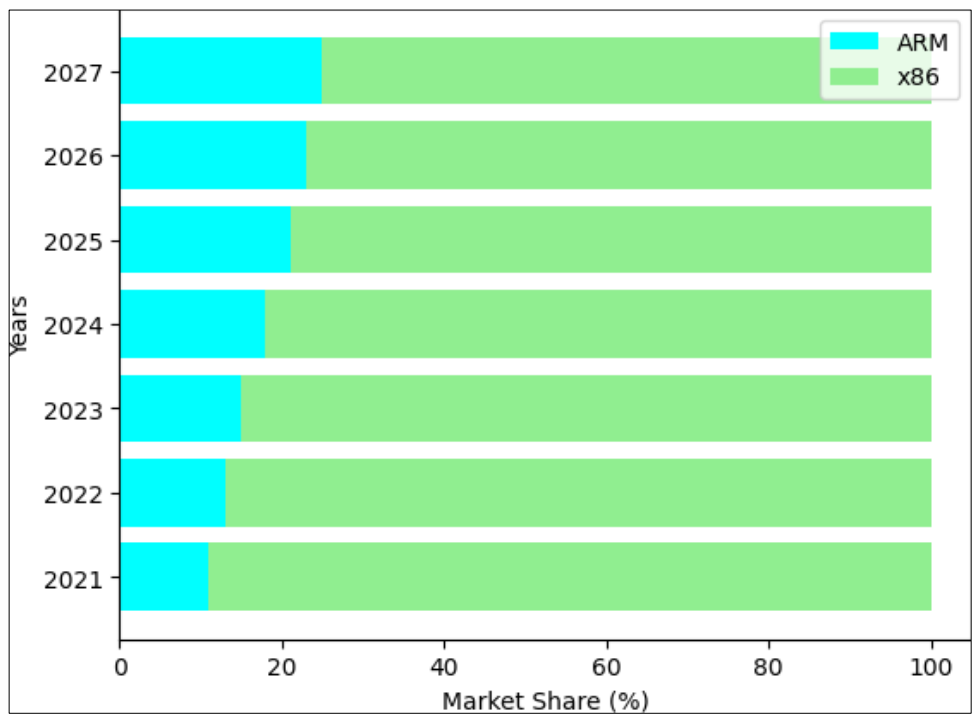

**Figure 3** Rapid growth of ARM laptops over the years

## 8. ARM stepped into Servers

A server is a computer that provides services to other computers or devices on a network, such as storing and sharing data, running applications, or hosting websites. A primary consideration regarding server power revolves around the energy usage and the ecological footprint of data centers, extensive facilities that accommodate numerous servers, and related equipment.

### 8.1. Early days

The AppliedMicro X-Gene 1 launched in 2014 is the first 64-bit ARM-based processor targeted at enterprise server workloads to reach commercial availability. While seeing limited deployment, it marked a significant milestone as the pioneer that validated ARM's potential in the data center.

AppliedMicro, now acquired by MACOM, began developing the X-Gene 1 around 2011 as one of the first efforts to bring ARM CPUs into servers and cloud infrastructure. At the time, ARM architectures were almost entirely confined to embedded and mobile use cases rather than server-grade applications.

The first-generation X-Gene design implemented high-performance ARM Cortex-A57 cores clocked up to 2.4GHz manufactured on a 40nm TSMC process.

Though conversion from the incumbent x86 standard was still slow, AppliedMicro's X-Gene 1 represented the pioneering first step towards opening the possibility of power-efficient ARM-based servers. It precedes more commercially impactful successors like Cavium ThunderX2, Amazon Graviton, and Ampere Altra. [33]

## 8.2. Data Centers

Data centers (facilities for processing, managing, and storing large amounts of data) consume significant electricity to power and cool the servers and support the network and backup systems. ARM-based servers use processors based on the ARM architecture, a popular RISC (Reduced Instruction Set Computing) design that efficiently handles complex tasks. ARM-based servers offer several advantages over traditional x86-based server architectures regarding power consumption, performance per watt (ppw), and overall system efficiency.

Ampere Computing tested while AMD EPYC 7742 delivers optimal ppw at 100%, their Altra Q80-33 exhibits notable ppw enhancements of 130%, 143%, 149%, and 118% in SPECrate 2017, Memcached, NGNIX, and Media Encoding, respectively. These servers are more affordable, customizable, and scalable than x86-based servers. ARM's licensing model allows other companies to create custom chips based on their architecture without starting from scratch, leading to rapid innovation and a diverse ecosystem. ARM server processors are highly scalable and capable of handling many workloads, from lightweight web servers to high-performance computing applications. So, ARM-based servers are ideal for cloud environments where conserving energy, reducing costs, and increasing flexibility is crucial. [34]

## 8.3. The Growing Market

The market for ARM server chips is growing rapidly as more and more cloud service providers, data center operators, and enterprise customers are adopting ARM-based servers for their computing needs. According to market research reports, ARM server chips are expected to take a significant share of the server market in the coming years, especially in cloud-native, edge, and high-performance computing workloads. Some of the key players who have defeated the x86 are discussed below:

### 8.3.1. AWS Graviton

AWS Graviton is a processor family developed by the Annapurna Labs subsidiary of AWS (Amazon Web Services) to deliver the best cost-effective cloud workload for Amazon EC2. The AWS Graviton processor is based on the ARM architecture and is compatible with the latest ARMv9 instruction set. They offer lower power consumption, higher performance, and higher price/performance than x86-based comparisons.

The first generation of the AWS Graviton processor launched in 2018 had 16 Cortex A72 cores operating at 2.3 GHz. The second generation of AWS Graviton processors, launched in 2019, had 64 Neoverse N1 cores at 2.5 GHz, and the third generation of AWS Graviton processors, launched in 2021, included 128 Neoverse N2 cores at 3.2 GHz. It offers 25% better computing performance than Graviton2 and improved memory bandwidth, cache size, and encryption acceleration. [35]

### 8.3.2. Ampere Altra

Ampere Altra is a server processor developed by Ampere Computing, founded by Renee James, former president of Intel. It is based on ARM architecture and is compatible with the latest ARMv9 instruction set. It is built on 7nm process technology and has 80 CPU cores that can run up to 3.3GHz. It has 8 DDR4 channels and 128 PCIe 4.0 lanes. This industry's first 80-core server processor provides high performance, scalability, and efficiency for cloud and edge computing workloads.

Ampere Altra was launched in March 2020 and adopted by several systems and cloud service providers such as Gigabyte, Lenovo, Oracle, Microsoft, and Tencent. It has a single-thread core, meaning that each core can perform a task at a time, which provides a more consistent and predictable performance than a multi-thread core that can share functions between multiple threads. Ampere Computing announced the next generation of the Ampere Altra, named

Ampere Altra Max, which supports up to 128 cores and DDR5 and PCIe 5.0. This was introduced in 2021 with further improvement in performance and efficiency. [36]

### 8.3.3. Yitian 710

The Alibaba Yitian 710 is a processor for servers designed by T-Head, the Alibaba Group's chip development wing. It is based on ARM architecture and compatible with the latest ARMv9 instruction set. It is built using 5nm process technology and has 128 CPU cores capable of running at 3.2GHz. This chip has 8 DDR5 channels and 96 PCIe 5.0 lanes. Yitian 710 is China's first ARM server processor aimed at improving Alibaba Cloud's computing performance and energy efficiency for native cloud workloads.

It was released at the Apsara Conference in October 2021, and Alibaba Cloud also announced the development of its server, Panjiu, powered by Yitian 710. Alibaba Cloud plans to use the Panjiu/Yitian combination to support current and future business in the Alibaba Group ecosystem and offer next-generation computing services to global customers. [37]

### 8.3.4. Azure Cobalt 100

Azure Cobalt 100 is a new server processor designed by Microsoft for Cloud Services. It is based on ARM architecture and is compatible with the latest ARMv9 instruction set. It is built on 5nm process technology and has 128 CPU cores with a performance of 3.2GHz. It also has 96 PCIe 5.0 lanes and 12 DDR5 channels, providing high memory and I/O bandwidth.

The Azure Cobalt 100 was announced at Microsoft Ignite 2023, revealing the first AI-powered accelerator series, Azure Maia. Microsoft also announced the next generation of the Azure Cobalt 200, which supports DDR6 and PCIe 6.0 with up to 256 cores. The Azure Cobalt 200 is expected to be available in 2025, further improving the performance and efficiency of Microsoft's cloud-native processors. [38]

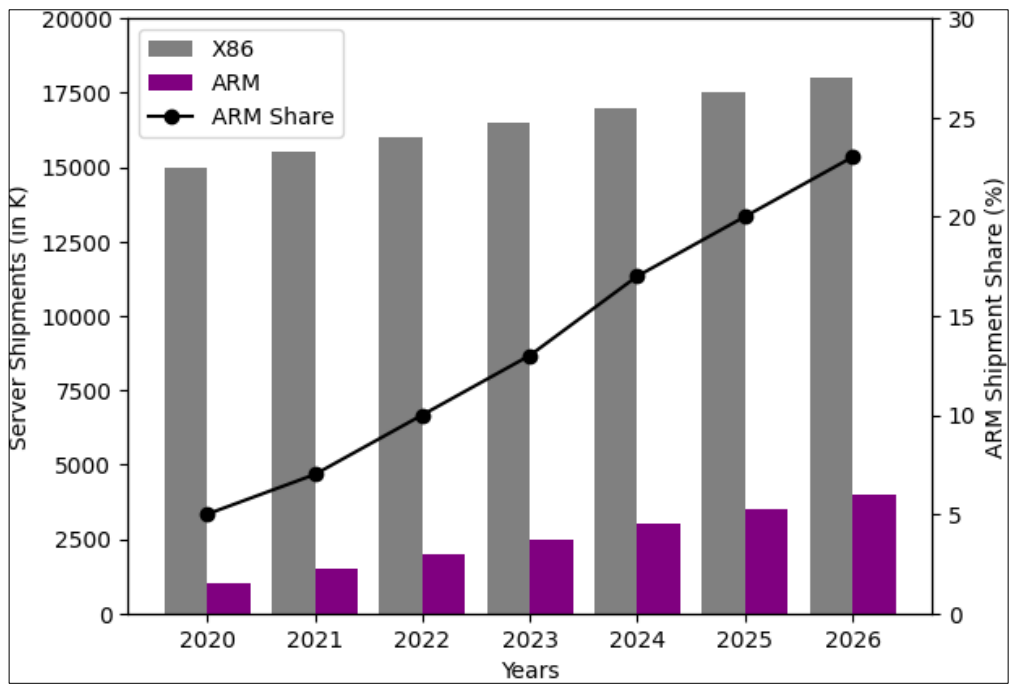

**Figure 4** Rapid growth of ARM-based servers over the years

More server-class chips are contributing to this ARM ecosystem. Huawei Kunpeng 920 is a server processor developed by Huawei and launched in 2019 for its cloud services. Nvidia Grace is a high-performance computing (HPC) processor developed by Nvidia for AI and scientific applications, released in 2023. Both chips are based on the ARM architecture and are compatible with the latest ARMv9 instruction set. Qualcomm also introduced the Centriq lineup, especially for data centers, back in 2017. Two other prominent examples are Maple and Cypress from Google. They are expected to be available in 2025 and power various Google services. [39]

Gartner reports that ARM server shipments have grown gradually, reaching approximately 77000 machines in 2020, 252100 machines in 2021, and an estimated 540400 machines in 2022. Projections indicate a significant increase to 934600 machines in 2023, 1.71 million in 2024, 2.54 million in 2025, and 3.2 million in 2026.

The success of ARM in the server realm hinges on the endorsement of hyperscalers (large-scale data centers) and cloud builders utilizing ARM for their workloads. The critical factor is whether customers will favor more affordable computing options from their cloud providers, provided that either internally developed or Ampere Computing's ARM server chips consistently offer better value. Gartner notes that ARM server sales surpassed $1 billion in a quarter for the first time in Q3 2022. [40]

### 8.4. Supercomputers

A supercomputer is a computer with very high performance, speed, and memory compared to a regular computer. Supercomputers are used for solving complex scientific and engineering problems that require a lot of computing power, such as weather forecasting, climate modelling, drug discovery, and artificial intelligence. They are usually composed of thousands or millions of processors that work together in parallel to perform calculations.

Fugaku is a supercomputer at the RIKEN Center for Computational Science in Kobe, Japan. It was jointly developed by RIKEN and Fujitsu in 2014 and started operation in 2020. It is named after an alternative name for Mount Fuji. Fugaku is the first supercomputer based on the ARM architecture, a popular design for mobile devices and power-efficient servers. Fugaku uses 158,976 Fujitsu A64FX processors, each with 48 cores, and supports the Scalable Vector Extensions. [41]

Fugaku became the fastest supercomputer in the world in June 2020, according to the TOP500 list, which ranks the performance of supercomputers using the High-Performance Linpack (HPL) benchmark. It achieved a speed of 415.5 petaflops, nearly three times faster than the second-ranked Summit system in the US. Fugaku also topped three other major supercomputer rankings, namely the HPCG, the Graph500, and the HPL-AI, making it the first supercomputer to achieve this feat. It reached the milestone of one exaflop, or one quintillion ($10^{18}$) floating-point operations per second, making it the world's fastest supercomputer. The Frontier system surpassed Fugaku in the US in May 2022, but it remains the second-fastest supercomputer in the world. [42] [43]

The significance of using the ARM chip in Fugaku is that it demonstrates the potential of the ARM architecture for high-performance computing (HPC), which the x86 architecture has traditionally dominated. ARM chips offer several advantages over the x86 chips, such as low energy consumption, higher watt performance, greater scalability, and customization. The ARM chip also allows Fugaku to run a wide range of applications, from cloud natives to artificial intelligence, with high efficiency and flexibility. Fugaku is also open-source friendly and contributes to the innovation and growth of the ARM ecosystem. This is a remarkable achievement in science and technology for Japan and a valuable asset for the global research community.

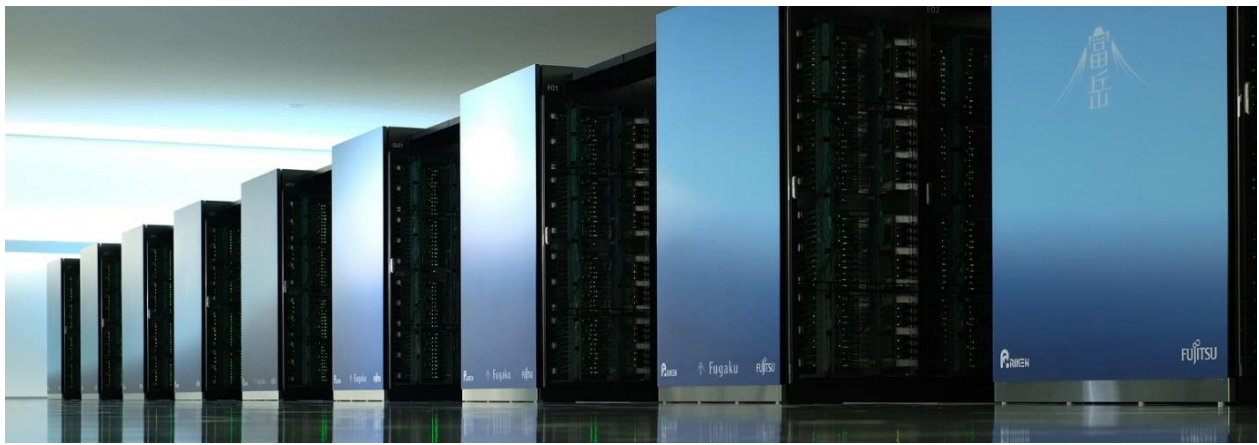

**Figure 5** Supercomputer Fugaku (富岳**)**

**Image credit:** RIKEN Center for Computational Science

## 9. Material and methods

Our study shows how much ISA influences the processor's overall performance, technically the whole device. The difference between Instruction Set Architectures (ISA) affects how we approach problem-solving. It is like using different methods to solve a mathematical problem with the same result. This difference in ISA affects various factors, such as power, performance, memory consumption, heat generation, etc., by determining the number of instruction cycles required for a task.

ARM processors use reduced instruction set computing (RISC) to simplify instructions and reduce the number of clock cycles required. This makes ARM chips more efficient, smaller in size, and ideal for portable devices, as they consume less power due to fewer transistors on the circuit board.

On the other hand, x86 processors use Complex Instruction Set Computing (CISC), which prioritizes maximum output but uses more clock cycles. This leads to more transistors, increased heat generation, and higher power consumption, as it doesn't prioritize power efficiency.

### 9.1. Test Subjects

We chose two laptops from Lenovo for our tests and two phones, one from ASUS and another from Oneplus. We had skipped the Microsoft Surface Pro X as it was too outdated to compete with modern chips. Then, we ran various experiments on these devices to differentiate actual power consumption and performance ratio in practical scenarios.

### 9.1.1. Laptops

Lenovo Thinkpad X13s Gen 1

It is an ARM laptop released in 2022 with the latest Qualcomm Snapdragon 8cx Gen 3 chip, specially built for computing platforms.

Lenovo Thinkpad X13 Gen 4

This is a x86-64 laptop from 2023 with Intel core i7-1355U processor. Here, the processor is marked as 'U,' which means it consumes ultra-low power. We'll study further from the test results.

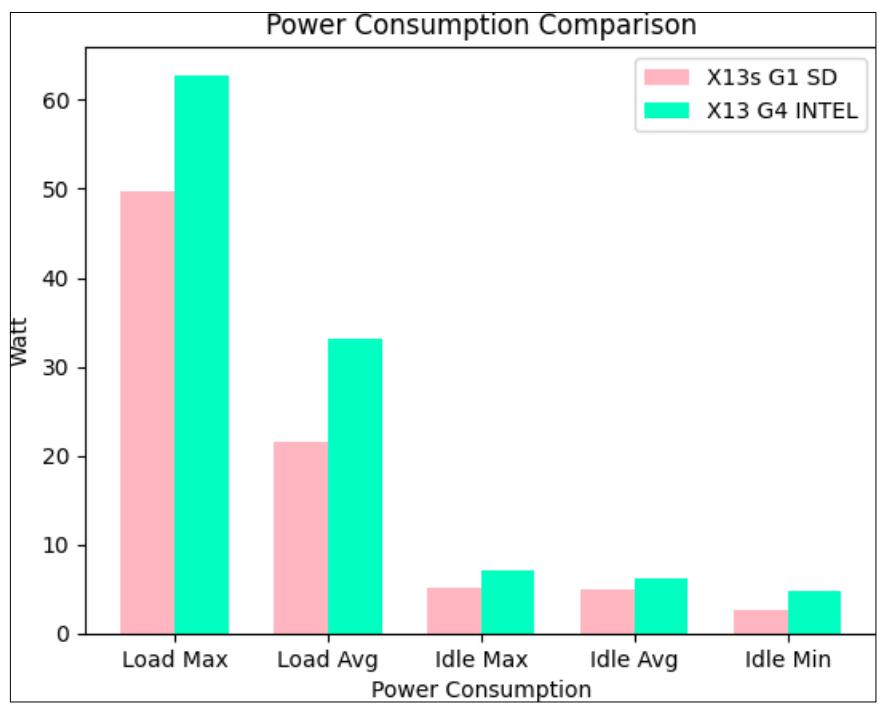

**Figure 6** Total power consumption over various conditions

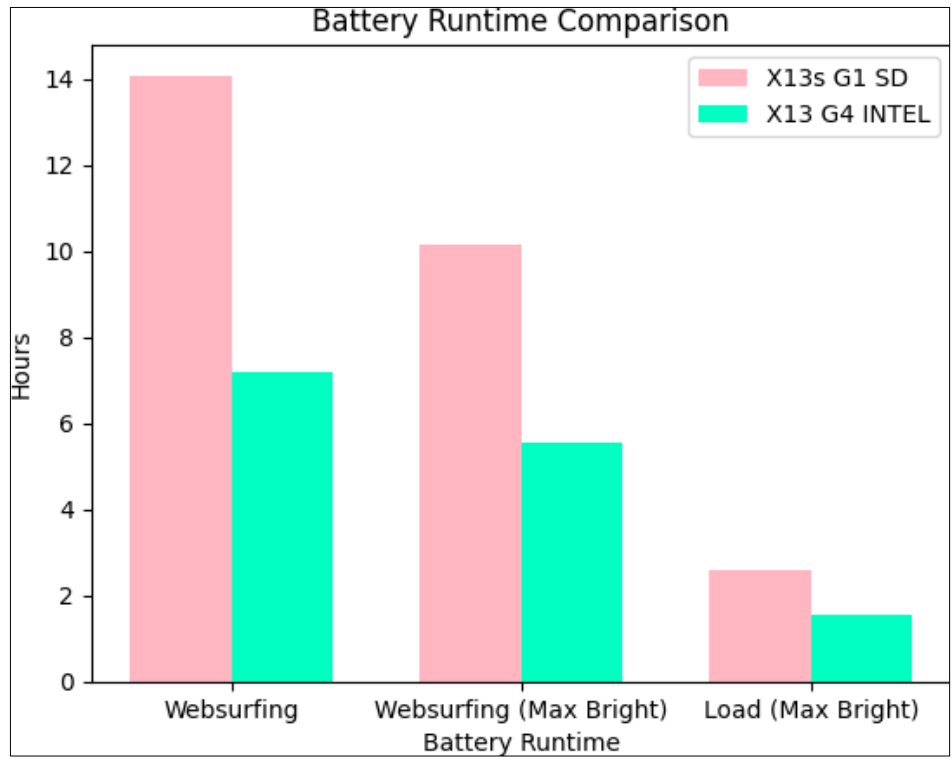

**Figure 7** Total runtime over various loads

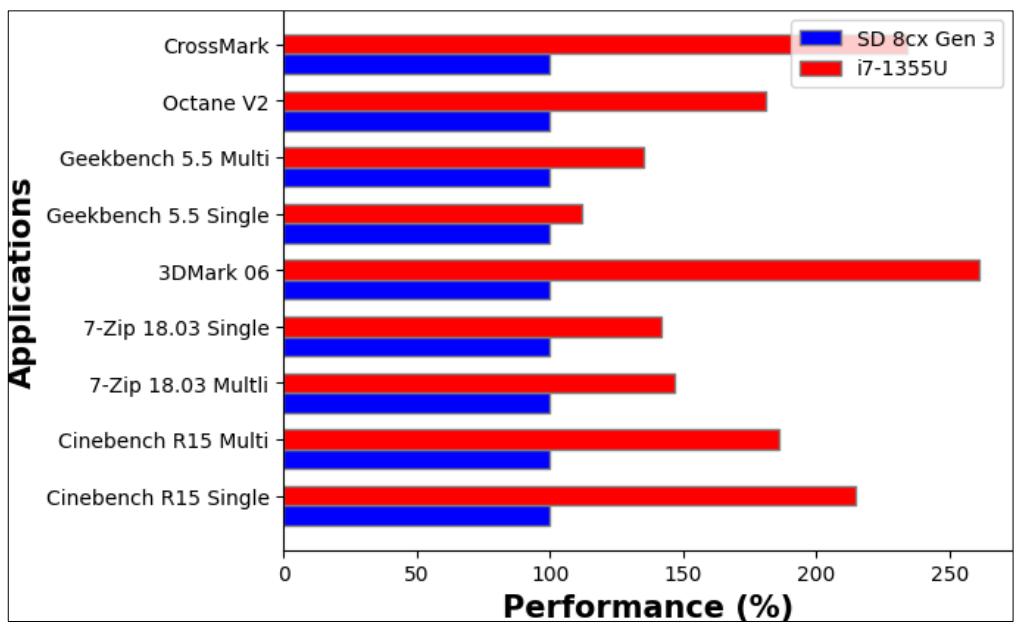

**Figure 8** Benchmarks between both chips

### 9.1.2. Smartphones

Asus Zenfone 2

We had introduced earlier.

Oneplus One

It is one of the most hyped phones from 2014, with the best price-to-performance ratio, containing Qualcomm's flagship ARM chip, Snapdragon 801. It has a clock speed of up to 2.5 GHz and is paired with an Adreno 330 GPU.

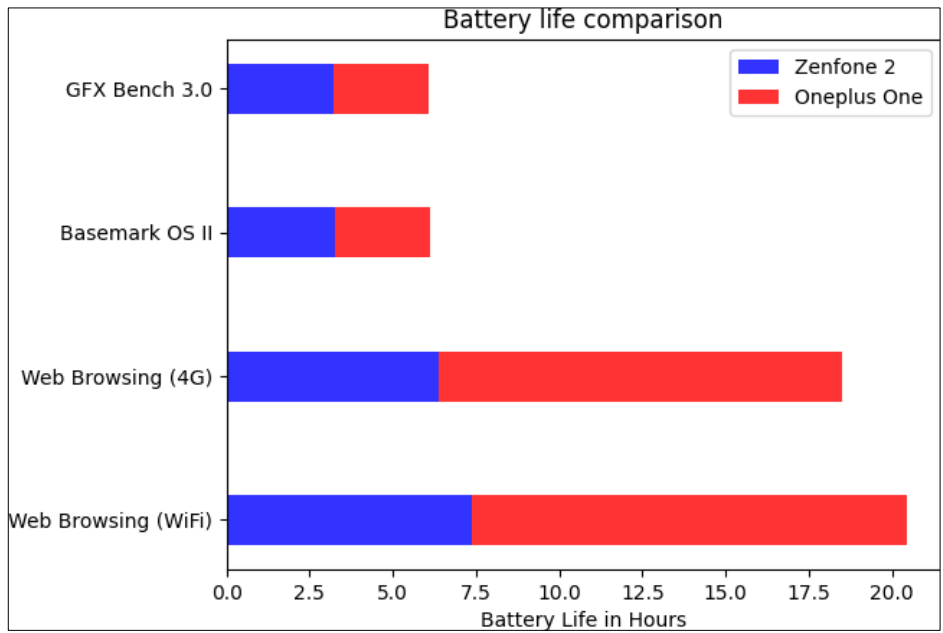

**Figure 9** Battery life in everyday usage and benchmarking

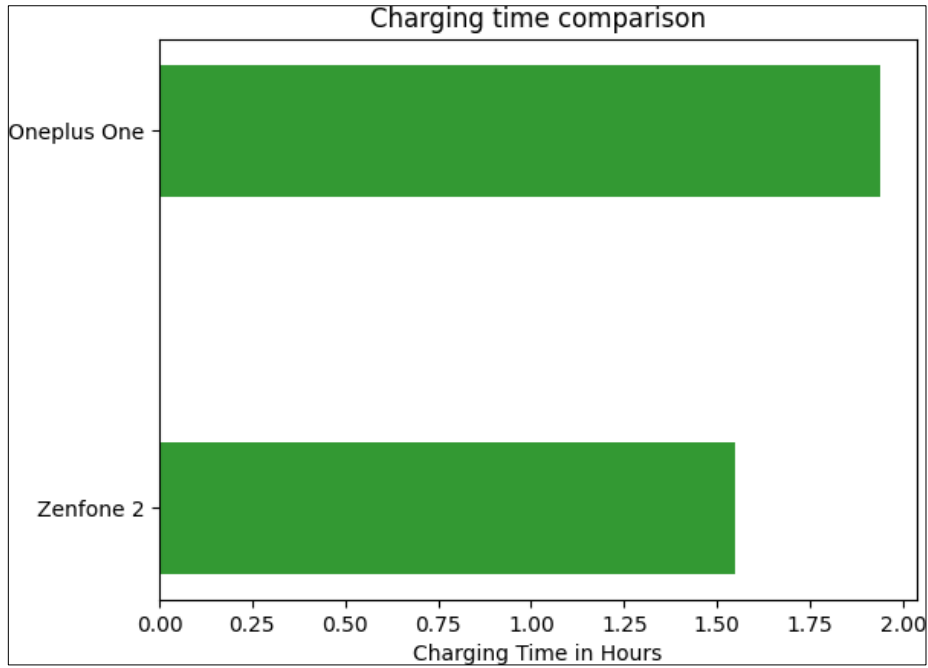

**Figure 10** Total time to charge from 0–100%

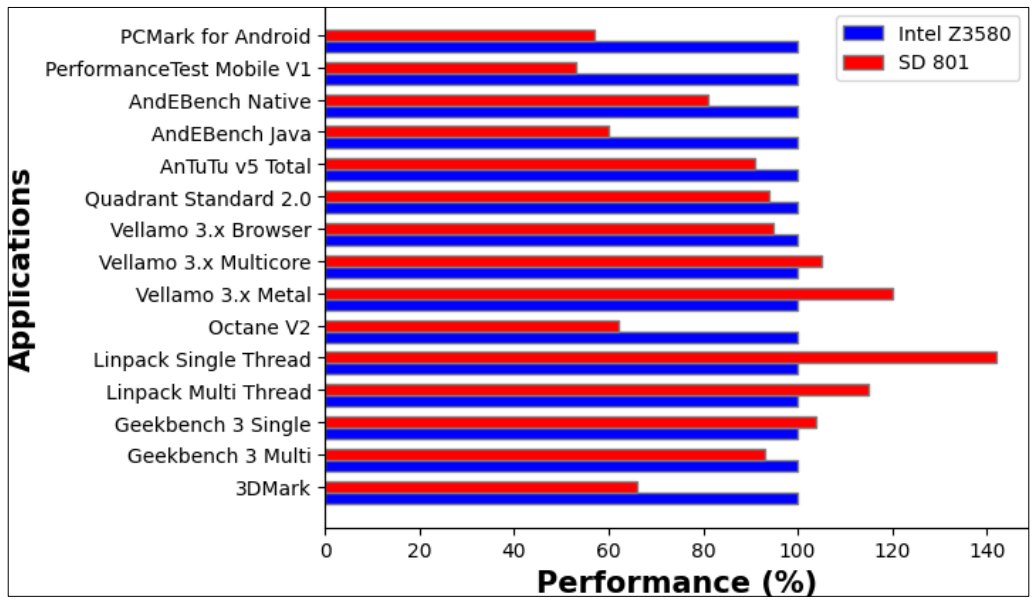

**Figure 11** Benchmarks between both chips

## 10. Results and Discussion

The results were more accurate than our expectations. As we initially said, x86 chips are more power-hungry than ARM. It was proved again in the battery runtime test from both laptops and smartphones. However, in the benchmark tests, x86 was the clear winner for its capability of solving complex problems efficiently.

### 10.1. Laptops

There is a big difference in the CPU cooling mechanism of both laptops. As X13s is optimized for heating issues, it generates significantly less heat that doesn't require a cooler. This is the main reason for the around double runtime compared to the X13 with the x86 chip, where a significant battery charge went off, causing heat and powering the cooler. However, i7 performed better than 8cx in every benchmark test, consuming more power. Eliminating the fan saves on system power drain in addition to the efficiency gains at the processor architecture level.

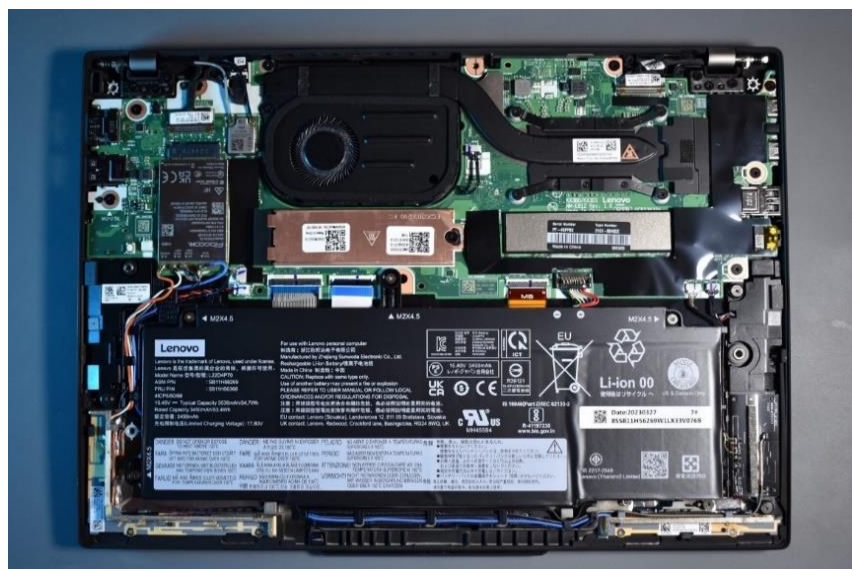

**Figure 12** X13 with a standard CPU cooler

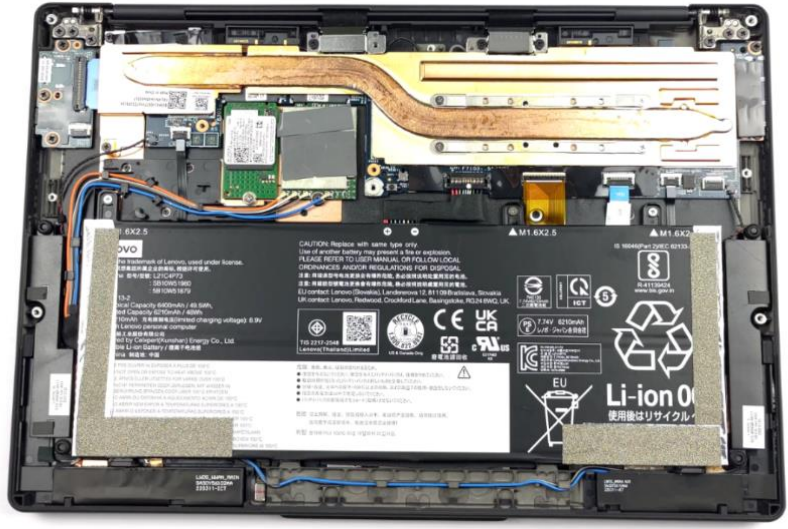

**Figure 13** X13s with no CPU cooler

TDP or Thermal Design Power refers to the maximum amount of heat generated by a chip that the cooling system in a computer is designed to dissipate under sustained workloads. A lower TDP typically means a cooler and more power-efficient processor.

The Qualcomm Snapdragon 8cx Gen 3 used in fanless laptops has an exceptionally low TDP of around 7 Watts owing to its highly energy-efficient ARM-based design and 5nm manufacturing process. This low TDP enables passive cooling through heat spreaders and pipes that quietly dissipate heat to the chassis instead of needing bulky fans. [44]

In contrast, the Intel Core i7-1355U is designed primarily for high performance and is manufactured on a thicker 14nm process. It has a much higher 15W TDP, generating significantly more heat under load. This requires an active cooling fan solution to prevent overheating in slim laptop bodies. The fan adds moving parts, noise, and higher power consumption. [45]

### 10.2. Smartphones

While the Asus Zenfone 2 and OnePlus One offer flagship specs, Zenfone 2's x86 Intel processor exacts a significant energy efficiency cost despite supplying faster performance. OnePlus One's Snapdragon 801 ARM chip had around 60% longer battery life but 20% slower charging speeds. Benchmark results reveal that the Atom Z3580 chip got nearly 2X higher scores in some cases, where the score of 801 was balanced. Ultimately, while the OnePlus One delivers a more well-rounded experience, blending performance and battery longevity, the Zenfone 2 prioritizes raw processing power for those willing to carry a battery pack to offset its energy hunger with fast charging.

These divergences mirror ARM and x86's philosophical clash between efficiency and performance, whatever the devices are.

## 11. Conclusion

The relationship between physics and processor heat stems from the conversion of electrical energy into computer operations, generating heat as per thermodynamic laws. RISC processors typically produce less heat than CISC processors because they utilize more straightforward instruction sets and streamlined microarchitectures. The streamlined microarchitectures exemplified by processors like ARM achieve enhanced efficiency through a combination of fundamental design principles. This includes using shorter pipelines with fewer stages, reducing the number of execution units and control logic, and simplifying data paths between components. With fewer transistors in action, heat production is lessened. Discarding complicated mechanisms in favor of a clean, optimized design allows RISC microarchitectures to function with remarkable clock speeds while maintaining lower power draw and thermal output. Additionally, these processors can make cloud services more accessible and cost-effective for a broader range of users, especially in developing countries where energy supplies and infrastructure are often limited or unreliable.

In simple terms, ARM is becoming a leader in sustainable and efficient computer architecture, emphasizing energy efficiency. On the other hand, x86 is better suited for scenarios that prioritize speed. This suggests a future where ARM and x86 play essential roles in different areas of computing. ARM's energy-efficient architecture will continue to be critical in addressing environmental concerns and enabling sustainability in computing. At the same time, x86 will remain a key player in high-performance computing (HPC) and other speed-sensitive applications.

Overall, the rise of ARM technologies signifies a growing trend toward computer systems designed to function effectively and ecologically. This forward-thinking approach ensures efficient performance and contributes to a global warming-conscious computing industry. As we move forward, this balance between efficiency and speed will shape the future of computer architecture, benefiting society.

## Compliance with ethical standards

### Disclosure of conflict of interest

No conflict of interest is to be disclosed.